\documentclass{elsart}

\usepackage{natbib}

 \usepackage{graphics}
 \usepackage{graphicx}
 \usepackage{epsfig}

\usepackage{amssymb}

\def\apj{ApJ }%
\def\apjs{ApJ Suppl. }%
\def\aap{A\&A }%
\def\solphys{Sol.~Phys. }%
\def\ssr{Sp.~Sci.~Rev. }%
\def\nat{Nature }%
\def\grl{Geophys.~Res.~Lett. }%
\def\jgr{J.~Geophys.~Res. }%
\def\memsai{Mem.~Soc.~Astron.~It. }%
\def\asr{Adv.~Space~Res. }%

\begin{document}
\begin{frontmatter}

\title{Reconstruction of solar irradiance using the Group sunspot number}

\author{L. Balmaceda, N.~A. Krivova and S.~K. Solanki}{}

\address{Max-Planck-Insitut f\"{u}r Sonnensystemforschung, \\
        Max-Planck-Str. 2, 37191 Katlenburg-Lindau, Germany \\
         e-mail: balmaceda@mps.mpg.de}

\begin{abstract}
We present a reconstruction of total solar irradiance since 1610 to the present
based on variations of the surface distribution of the solar magnetic field.
The latter is calculated from the historical record of the Group sunspot
number using a simple but consistent physical model. Our model successfully
reproduces three independent data sets: total solar irradiance measurements
available since 1978, total photospheric magnetic flux from 1974 and the open
magnetic flux since 1868 (as empirically reconstructed from the geomagnetic
aa-index). The model predicts an increase in the total solar irradiance since
the Maunder Minimum of about 1.3~\rm{Wm$^{-2}$}.
\end{abstract}


\end{frontmatter}

\section{Introduction}
\label{intro}

Total solar irradiance changes by about $0.1\%$ between solar activity
maximum and minimum. Accurate measurements of this quantity are only
available since 1978 \citep{froehlich06} and do not provide information on
longer-term secular trends. In order to reliably evaluate the Sun's role in
recent global climate change, longer time series are, however, needed. They
can only be assessed with the help of suitable models. The most successful
models are those attributing irradiance variations on timescales longer than
a day to the evolution of the Sun's surface magnetic field
\citep{solanki05,krivova05}. Such models explain more than 90$\%$ of all
observed changes in the total and spectral irradiance at these timescales
\citep{krivova03,wenzler04,wenzler05,wenzler06,krivova06}. The continuously
evolving distribution of the solar magnetic field on the surface is described
in these models by recourse to magnetograms, which are only available since
1974. For a longer term reconstruction, another proxy of solar magnetic
activity has to be employed. The available historical proxies of solar
activity, such as the Group and Zurich sunspot numbers, sunspot, facular or
Ca II plage areas mainly describe the evolution of the larger magnetic
features, such as sunspots or faculae, but do not provide any direct
information about the weaker features. Therefore, whereas the reconstruction
of the cyclic component of the irradiance variation is typically not a
problem, evaluation of the amount of the secular change is not
straightforward \citep[see][and references therein]{solanki04a}.
\citet{solanki00,solanki02} proposed a simple physical mechanism which can
lead to such a secular trend in the magnetic flux based on the overlap of
consecutive activity cycles. Here, we use their model to reconstruct the
magnetic flux of the Sun back to 1610 from the Group sunspot number
\citep{Hoyt98} which is employed to reconstruct total solar irradiance for
the same period.

\section{Approach}
 \label{approach}

\subsection{Photospheric Magnetic Flux}
\label{magflux}
The basic assumption of our model is that the irradiance variations
are caused entirely by the evolution of the magnetic features on the solar
surface. As in the model of \citet{solanki02}, magnetic features on the Sun's
surface are divided into active regions (AR) and ephemeral regions (ER). The
flux emergence rate in AR, $\phi_{act}$, can be estimated from the Group
sunspot number since it serves as a good proxy for the fresh flux threading
the solar surface. The time evolution of the flux emerging in ephemeral
regions, $\phi_{eph}$, is more uncertain, however. Observations suggest that
the emergence rate in ER is related to that of AR, but the exact shape of
this relationship is not yet well established. They show that ER associated
with the new cycle start emerging at the solar surface before the
corresponding AR cycle begins and while magnetic features from the previous
cycle are still appearing \citep{harvey92,harvey93}. Thus, the ER cycle
length is extended with respect to that of AR. Therefore we prescribe a sine
approximation for the shape of the ER cycle, with its length being somewhat
stretched in time with respect to that of the corresponding AR cycle and with
its amplitude being proportional to the amplitude of the AR cycle
\citep[see][for details]{krivova07}.

Both active and ephemeral regions contribute to the open flux ($\phi_{open}$), which is
dragged by the coronal gas and reaches far into the heliosphere. Since this
flux is mainly unipolar, it decays slowly and lives much longer on the solar
surface (up to several years) than the AR and ER flux.

The extended length of the ER cycles and the long lifetime of the open flux
lead to an overlap between consecutive cycles such that some background
magnetic flux is present on the solar surface even at activity minima. The
amount of this flux changes with time due to variations in the length and the
amplitude of the magnetic activity cycle. This mechanism thus provides a
physical explanation for a secular change in the total photospheric magnetic
flux.

\subsection{Variations of the total solar irradiance}
\label{irra}

Following \citet{krivova03,wenzler05}, the solar
photosphere is divided into 5 atmospheric components: the quiet Sun, sunspot
umbrae and penumbrae, faculae and the network, denoted with subindices
\textit{q, u, p, f} and \textit{n}, respectively. The model consists of two
main ingredients: one which is temporally invariant and another that
introduces a variation with time.

The time-independent brightness of each component $F_{q,u,p,f,n}(\lambda)$,
with $\lambda$ being the wavelength, is calculated using the ATLAS9 code of
Kurucz from plane-parallel model atmospheres \citep[see][for a
description of the models]{unruh99}. Faculae and the network are described by
the same model atmosphere. All fluxes obtained in this way depend only on the
wavelength.

On the other hand, variability in time is due to the changing surface
distribution of the magnetic components. To describe this, we need to
determine which part of the solar surface is covered by each component at a
given time, i.e. the corresponding filling factors, $\alpha$. In case of
sunspots they are directly extracted from the sunspot area time series
available since 1874 \citep[see][]{balmaceda05}. Before that time, sunspot
areas are extrapolated by comparing them with the sunspot number. To
estimate the filling factors for umbrae and penumbrae separately, we use the
umbral to penumbral area ratio $\alpha_u/(\alpha_u+\alpha_p)=0.2$ as found by
\citet{wenzler06}. The filling factors of the other components are obtained
from the reconstructed solar magnetic flux.
The magnetic flux in faculae, $\phi_f$, can be obtained from
$\phi_f=\phi_{act}-\phi_u - \phi_p$, where $\phi_{act}$ is the flux in AR and
$\phi_{u},\phi_p$ represent the flux in sunspot umbrae and penumbrae,
respectively. Finally, the evolution of the network magnetic flux, $\phi_{n}$,
which is responsible for the secular change, is given by the sum of the
flux from ER, $\phi_{eph}$, and the open flux, $\phi_{open}$:
$\phi_{n}=\phi_{eph}+\phi_{open}$.

The final step is the conversion of the magnetic flux in faculae and the
network into the corresponding filling factors. For this, we follow the
conversion scheme described by \citet{fligge00} and \citet{krivova03}. The facular and
network filling factors increase linearly from 0 at $\phi=0$ to 1 at
$\phi_{sat}$. For magnetic flux larger than $\phi_{sat}$, the filling factor
remains unity. Following \citet{krivova03,wenzler04,wenzler05,wenzler06} we
use the value $\phi_{sat,f} = 300~\rm{G}$ for faculae while for the network we
employ $\phi_{sat,n}= 500~\rm{G}$ . This somewhat enhanced saturation level for
the network takes into account the fact that a significant amount of weak
magnetic flux in the network is lost due to the insufficient spatial
resolution and relatively high noise level of the magnetograms employed.

Finally, the solar radiative flux at a given wavelength, $\lambda$, can be
obtained by combining the fluxes from the 5 components:

\begin{eqnarray*}
\nonumber 
F(\lambda,t) & = & \alpha_q(t)F_q(\lambda) +\\
& + & \alpha_u(t)F_u(\lambda) +\\
& + & \alpha_p(t)F_p(\lambda) +\\
& + & (\alpha_f(t) + \alpha_n(t)) \cdot F_f(\lambda).
\end{eqnarray*}

Here, $\alpha_q(t)=1-\alpha_u(t)-\alpha_p(t)-\alpha_n(t)-\alpha_f(t)$. The
total solar irradiance is then obtained by integrating $F(\lambda,t)$ over all wavelengths.

\section{Results}
\label{results}

The reconstructed total magnetic flux for individual Carrington rotations is
compared to data obtained at different observatories, WSO, NSO KP and MWO, in
Fig. 1a. The model reproduces both the amplitude and the length of the solar
cycle in the observed magnetic flux.  The evolution of the
magnetic flux in active and ephemeral regions, as well as of the open
and total magnetic flux is shown in Fig. 1b. The effect of the spatial
resolution on the detection of small ER noted by \citet{krivova04} is taken
into account by considering the quantity $\phi_{tot}=\phi_{act}+0.4 \cdot
\phi_{eph}+\phi_{open}$. Since ER cycles overlap, the flux in these regions
varies only by a factor of 2 over a cycle, which is much smaller than the
variation in the AR cycles. The ER flux is comparable to that of AR during
activity maxima while it clearly dominates during minima, in agreement with
\citet{harvey93} and \citet{krivova04}.

Our model also reproduces the secular increase in the open flux during the
last century found by \citet{lockwood99}. This is illustrated in Fig.~2,
where the modelled open flux is compared to the reconstruction based on the
aa-index by these authors.

Figure 3a shows the comparison between the PMOD composite of TSI
\citep[][grey solid line]{froehlich06} and the solar irradiance reconstructed
from the Group sunspot number (dotted line). In order to facilitate
comparison, we plot 3-month running means. Note, however, that individual
dips due to sunspots are also reproduced if daily time series is considered.
Both the amplitude and the phase of the variations are reasonably
reconstructed. Of course, the coarse magnetic flux model we use is rather
simple to reproduce all details of the irradiance variations. For example,
the sine approximation of the cycle shape does not produce the double-peak
structure of cycle 23, as was observed.

The daily values of the reconstructed solar irradiance since 1610 are shown
in Fig. 3b. The 11-yr running mean is indicated by the grey solid line. As
mentioned before, for the period prior to 1874 when no sunspot
area measurements are available, they are obtained by extrapolating sunspot
number back to 1611. For the period prior to 1753 only monthly values of the
Group sunspot number can be used. After 1753 daily values are available
although prior to approximately 1850 there are many gaps in the data.
Sampling becomes more regular after 1818. Our model predicts an increase in
the solar irradiance since the end of the Maunder Minimum (i.e., 1700) till
the present (average over about 30 years) of about $0.095\%$ or
1.3~\rm{Wm$^{-2}$}.

\section{Conclusion}

We have reconstructed total solar irradiance back to 1610. The cyclic
variation of ER was assumed to be related to the properties of the
corresponding AR cycle, whose variation can be estimated from the Group
sunspot number \citep{solanki02}. The secular change in the total magnetic
flux of the Sun and, therefore, in the irradiance is caused by the overlap of
the consecutive ER cycles. The predicted secular change since 1700 is about
1.3~\rm{Wm$^{-2}$}. This value lies within the range suggested by other
recent reconstructions of solar irradiance \citep{foster04,wang05}, but is
significantly lower than the ones obtained in earlier investigations based on
stellar data ranging from 2 to 16~\rm{Wm$^{-2}$}
\citep[e.g.,][]{mendoza97,lean00a}. However, the stellar evidence for such a
change has been recently critized \citep{hall04,wright04,giampapa05} and the
magnitude of the increase in TSI obtained using these results might have been
overestimated.

\section{Acknowledgments}
This work was supported by the \textit{Deutsche Forschungsgemeinschaft, DFG} project
number SO 711/1-1.

\bibliographystyle{elsart-harv}

\begin{figure}
\begin{center}
\resizebox{\hsize}{!}{\includegraphics{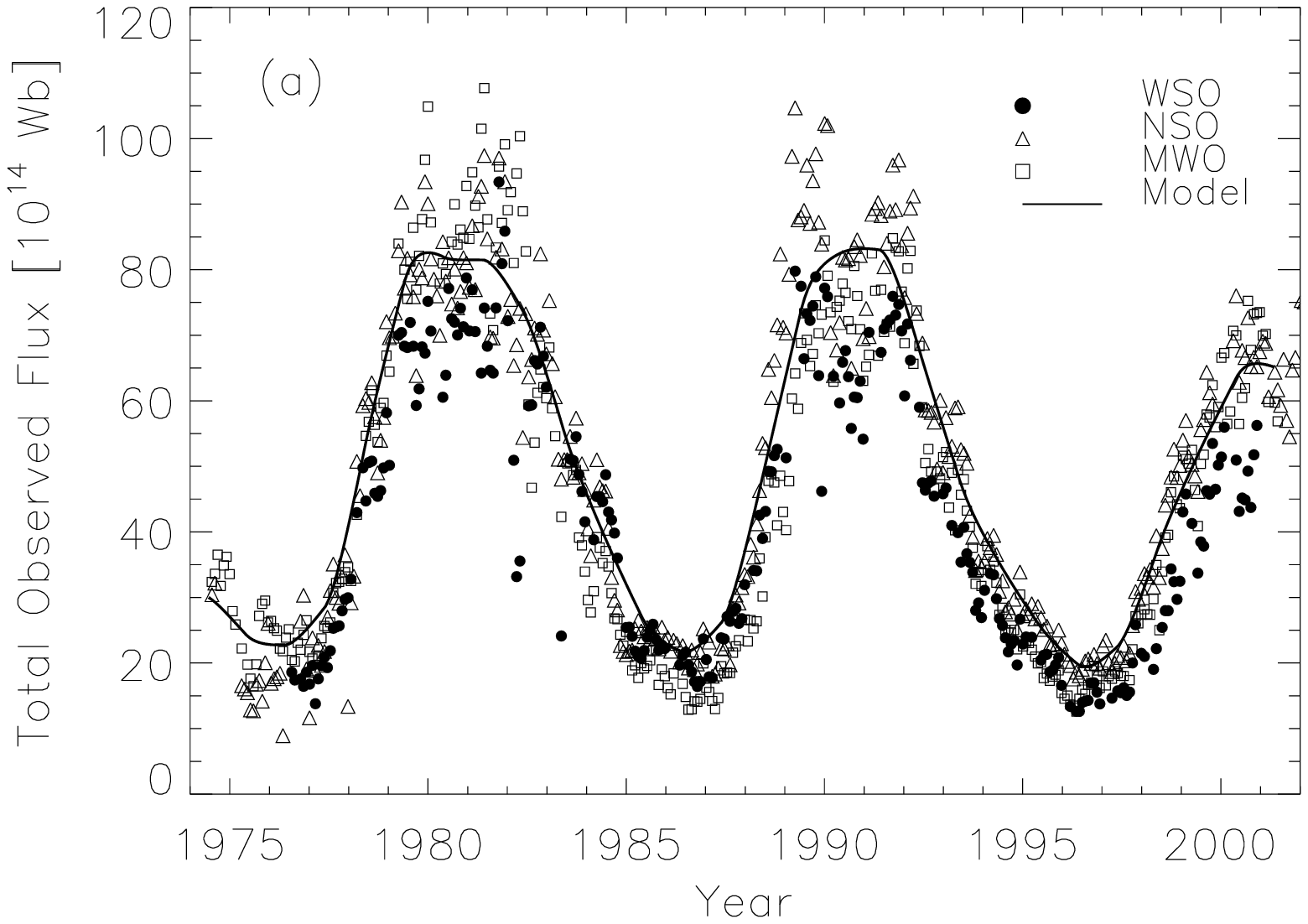}}
\resizebox{\hsize}{!}{\includegraphics{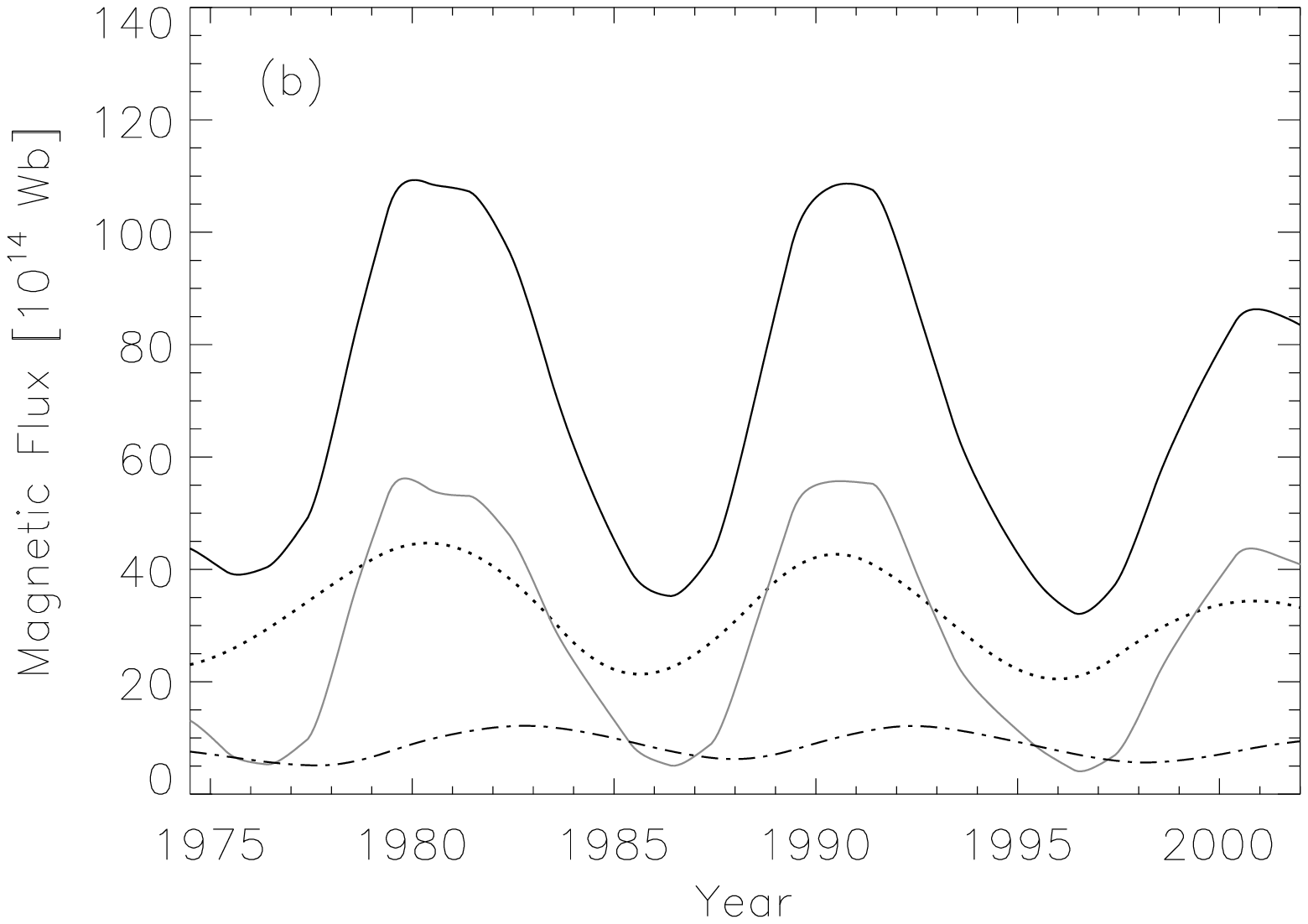}} \caption{\footnotesize
(a): Total magnetic flux for individual Carrington rotations between 1974 and
2002. Different symbols represent data from synoptic charts obtained by
different observatories \citep{arge02}. The solid line shows the magnetic
flux calculated from the Group sunspot number. (b): Reconstructed magnetic
flux in active regions (grey line), ephemeral regions (dotted), as well as
the open (dot-dashed) and total flux (black line).} \label{fig1}
\end{center}
\end{figure}

\begin{figure}
\begin{center}
\resizebox{\hsize}{!}{\includegraphics{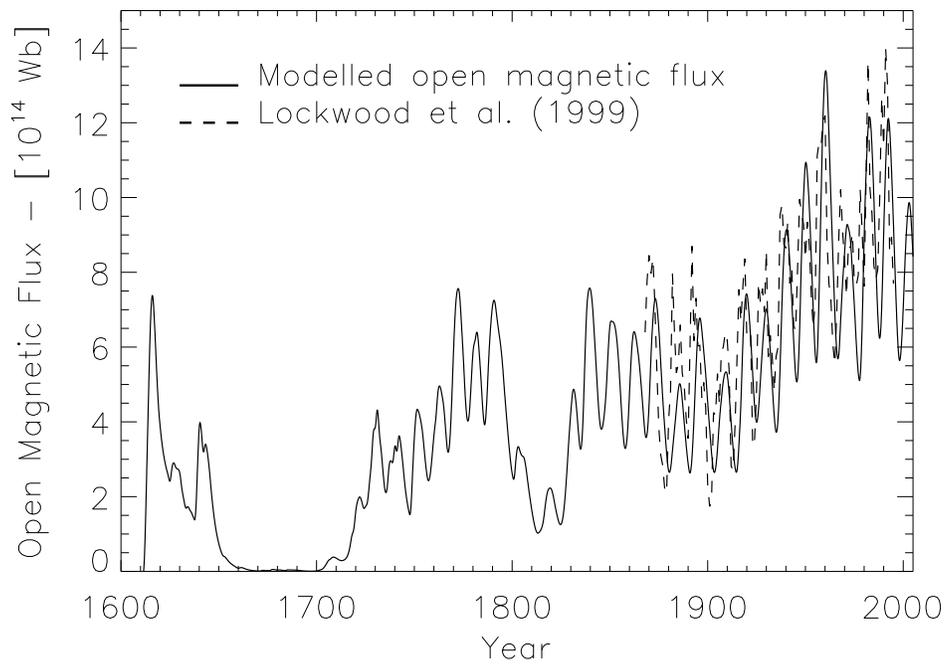}} \caption{\footnotesize The
calculated open flux (solid line) and the reconstruction based on the
geomagnetic \textit{aa}-index by \citet{lockwood99}.} \label{fig2}
\end{center}
\end{figure}

\begin{figure}
\begin{center}
\resizebox{\hsize}{!}{\includegraphics{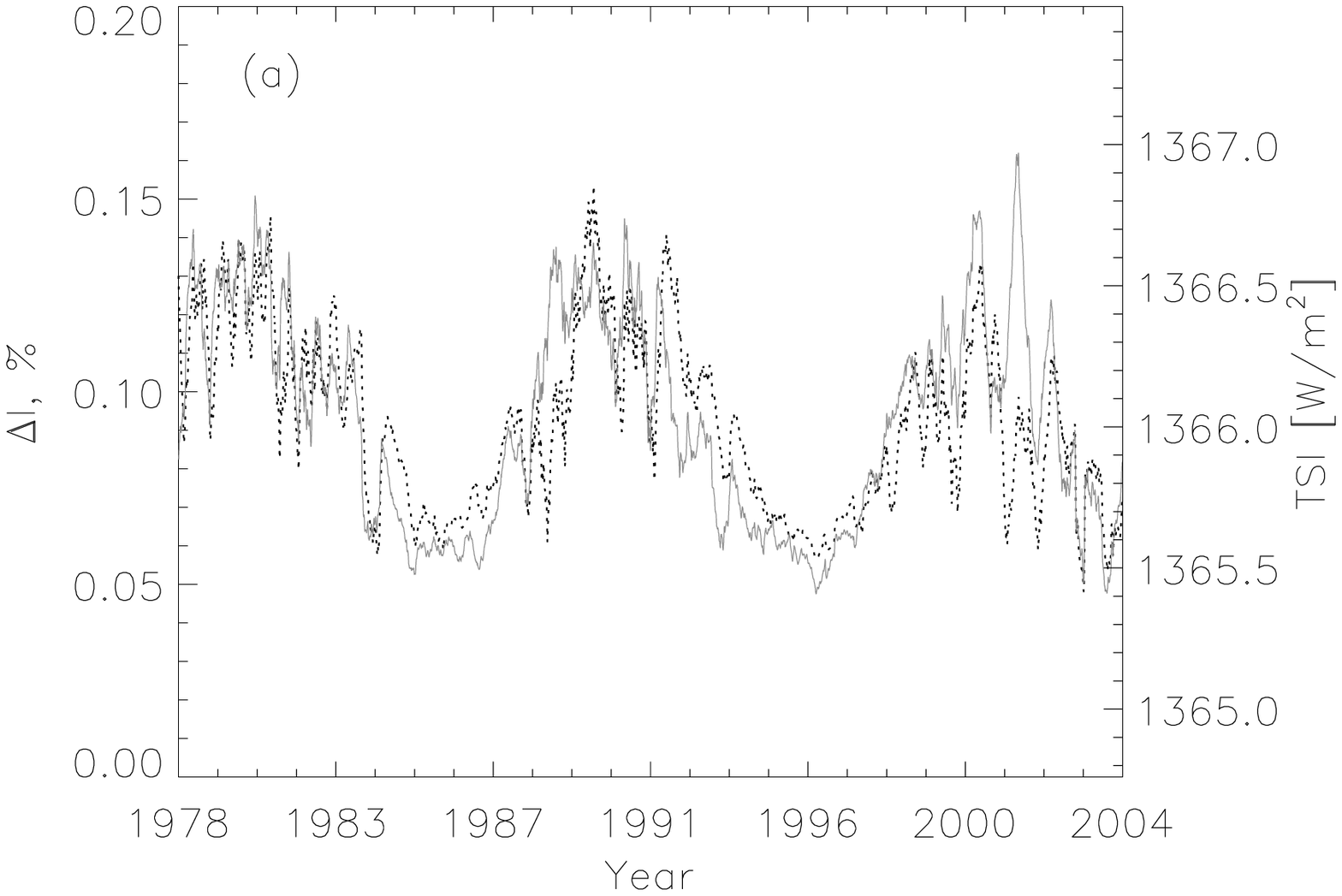}}
\resizebox{\hsize}{!}{\includegraphics{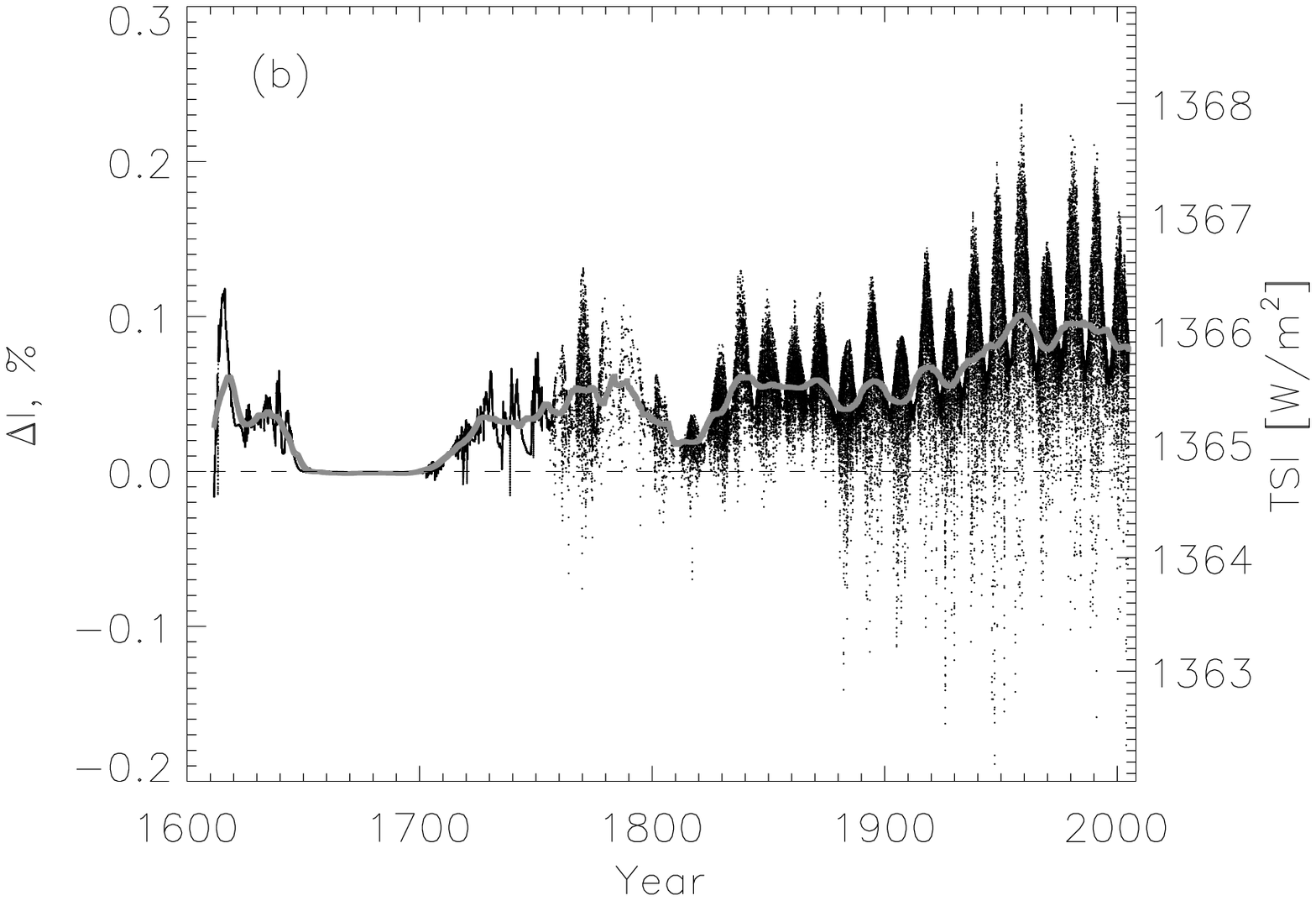}}
\caption{\footnotesize (a): 3-month running mean of the daily sampled TSI for
recent cycles: modelled (dots) and measured (solid line; PMOD composite). (b):
Variation of the total solar irradiance since 1610. The grey solid line
represents the 11-yr running mean of this variation.} \label{fig3}
\end{center}
\end{figure}

\end{document}